\documentclass[aps,prl,reprint,superscriptaddress,nofootinbib]{revtex4-1}

\usepackage{lmodern}
\usepackage[T1]{fontenc}
\usepackage[english,activeacute]{babel}
\usepackage{mathtools}
\usepackage{color} % color package

\usepackage{amssymb} %letras chulis
\usepackage{dsfont} % para escribir la matrix identidad chula

\usepackage{pbsi}

\usepackage[colorlinks,linkcolor=blue,citecolor=blue,urlcolor=blue]{hyperref}

\addto\extrasenglish{}
\addto\extrasenglish{}
\begin{document}

% New commands:

\newcommand{\sC}{\mbox{\tiny{C}}}
\newcommand{\sA}{\mbox{\tiny{A}}}
\newcommand{\sD}{\mbox{\tiny{D}}}

\newcommand{\sU}{\mbox{\tiny{U}}}
\newcommand{\sk}{\mbox{\tiny{k}}}
\newcommand{\sK}{\mbox{\tiny{K}}}
\newcommand{\sM}{\mbox{\tiny{M}}}
\newcommand{\sx}{\mbox{\tiny{x}}}
\newcommand{\sX}{\mbox{\tiny{X}}}
\newcommand{\sL}{\mbox{\tiny{L}}}

\newcommand{\mD}{\tiny{\mathcal{D}}}
\newcommand{\mA}{\tiny{\mathcal{A}}}

\newcommand{\sUP}{\mbox{\tiny{UP}}}
\newcommand{\sMP}{\mbox{\tiny{MP}}}
\newcommand{\sLP}{\mbox{\tiny{LP}}}

\newcommand{\sP}{\mbox{\tiny{P}}}

\newcommand{\sR}{\mbox{\tiny{R}}}

\newcommand{\Tr}{\mbox{Tr}}

\newcommand{\dd}{\mbox{d}}
\newcommand{\ii}{\mbox{i}}
\newcommand{\ddd}{\mbox{\scriptsize{d}}}
\newcommand{\iii}{\mbox{\scriptsize{i}}}

\newcommand{\QE}{\mbox{\tiny{QE}}}
\newcommand{\LL}{\mbox{\tiny{L}}}
\newcommand{\C}{\mbox{\tiny{C}}}

\newcommand{\LP}{\mbox{\tiny{LP}}}
\newcommand{\UP}{\mbox{\tiny{UP}}}

\newcommand{\JC}{\mbox{\tiny{JC}}}

\newcommand{\g}{\mbox{g}}
\newcommand{\e}{\mbox{e}}
\newcommand{\gmini}{\mbox{\scriptsize{g}}}
\newcommand{\emini}{\mbox{\scriptsize{e}}}
\newcommand{\eg}{\mbox{\scriptsize{eg}}}

\newcommand{\eff}{\mbox{\scriptsize{eff}}}

\newcommand{\intt}{\mbox{\scriptsize{int}}}

\newcommand{\HH}{\mbox{\tiny{H}}}
\newcommand{\VV}{\mbox{\tiny{V}}}

\newcommand{\II}{\mbox{\tiny{I}}}

\newcommand{\rr}{\mbox{\tiny{r}}} % radiative
\newcommand{\nr}{\mbox{\tiny{nr}}} % non-radiative

\newcommand{\dtQE}{\tilde{\Delta}_{\QE}} %
\newcommand{\dtC}{\tilde{\Delta}_{\C}} %

\newcommand{\Ed}{\boldsymbol{E}_{\rm D}}

\title{Organic Polaritons Enable Local Vibrations to Drive Long-Range Energy Transfer}

\author{R. S\'aez-Bl\'azquez}
\affiliation{Departamento de F\'isica Te\'orica de la Materia Condensada and Condensed Matter Physics Center (IFIMAC), Universidad Aut\'onoma de
    Madrid, E-28049 Madrid, Spain}

\author{J. Feist}
\affiliation{Departamento de F\'isica Te\'orica de la Materia Condensada and Condensed Matter Physics Center (IFIMAC), Universidad Aut\'onoma de
    Madrid, E-28049 Madrid, Spain}

\author{A. I. Fern\'andez-Dom\'inguez}
\affiliation{Departamento de F\'isica Te\'orica de la Materia Condensada and Condensed Matter Physics Center (IFIMAC), Universidad Aut\'onoma de
    Madrid, E-28049 Madrid, Spain}

\author{F. J. Garc\'ia-Vidal}
\email{fj.garcia@uam.es}
\affiliation{Departamento de F\'isica Te\'orica de la Materia Condensada and Condensed Matter Physics Center (IFIMAC), Universidad Aut\'onoma de
    Madrid, E-28049 Madrid, Spain}
\affiliation{Donostia International Physics Center
    (DIPC), E-20018 Donostia/San Sebasti\'an, Spain}

\begin{abstract}
Long-range energy transfer in organic molecules has been
experimentally obtained by strongly coupling their electronic 
excitations to a confined electromagnetic cavity
mode. Here, we shed light into the polariton-mediated mechanism
behind this process for different configurations: donor and
acceptor molecules either intermixed or physically separated.
We numerically address the phenomenon by means of Bloch-Redfield theory,
which allows us to reproduce the effect of complex vibrational
reservoirs characteristic of organic molecules. Our findings reveal
the key role played by the middle polariton as the non-local
intermediary in the transmission of excitations from donor to
acceptor molecules. We also provide analytical insight on the key physical magnitudes 
that helps to optimize the efficiency of the long-range energy transfer.
\end{abstract}

\maketitle

Energy transfer is crucial in the process of photosynthesis in
biological complexes~\cite{Grondelle2006,Scholes2011,OlayaCastro2011}. In recent
years, fundamental research on this phenomenon has also led to the
proposal of optoelectronic devices mimicking it for different
functionalities~\cite{Baldo2000,Hardin2009}. In these systems, the
exciton transmission from one organic molecule to another relies on their short-range
dipole-dipole interaction, often described by F\"orster theory~\cite{Forster1948}.
This restricts the efficiency of the process to
nanometric distances. However, by modifying the electromagnetic environment 
of the molecules, several experimental studies \cite{Andrew2004,Gotzinger2006} have shown  
that it is possible to extend the range of energy transfer to tens of nanometers. 

On the other hand, very recent experimental and theoretical works have demonstrated 
that when a collection of organic molecules is strongly coupled to an electromagnetic mode~\cite{Pockrand1982,Lidzey1998, Bellessa2004,
Dintinger2005,Torma2015}, 
their chemical~\cite{Hutchison2012,Galego2015,Herrera2016,Bennett2016,Flick2017} 
and material~\cite{Orgiu2015,Feist2015,Schachenmayer2015} properties can be tailored.
Regarding the energy transfer process, it has been also shown that it is feasible 
to enhance its efficiency by taking advantage of the phenomenon of collective strong coupling~\cite{Coles2014,Ebbesen2016}. 
This gives rise to spatially-extended hybrid light-matter states 
(i.e., polaritons), which can be utilized to extend 
the range of energy transfer to length scales comparable to the optical wavelength of the cavity 
mode (hundreds of nanometers)~\cite{Ebbesen2017,FJFeist2017,Joel2017,Genes2018}.

In this Rapid Communication, we present a theoretical description
of this non-local energy transfer ---beyond nearest neighbor
interactions--- taking place between two sets of organic molecules
(donors and acceptors) strongly coupled to a cavity mode. The use of the Bloch-Redfield theory allows us to
introduce the effect of vibrational reservoirs, which act as an effective vehicle for
the excitation transmission from donor to acceptor molecules. The
prevailing decay path turns out to involve the so-called middle
polaritons, whose mixed composition of donor and acceptor states
enables the energy transfer.
\begin{figure}[!t]
    \centering
    \includegraphics[width=0.95\columnwidth]{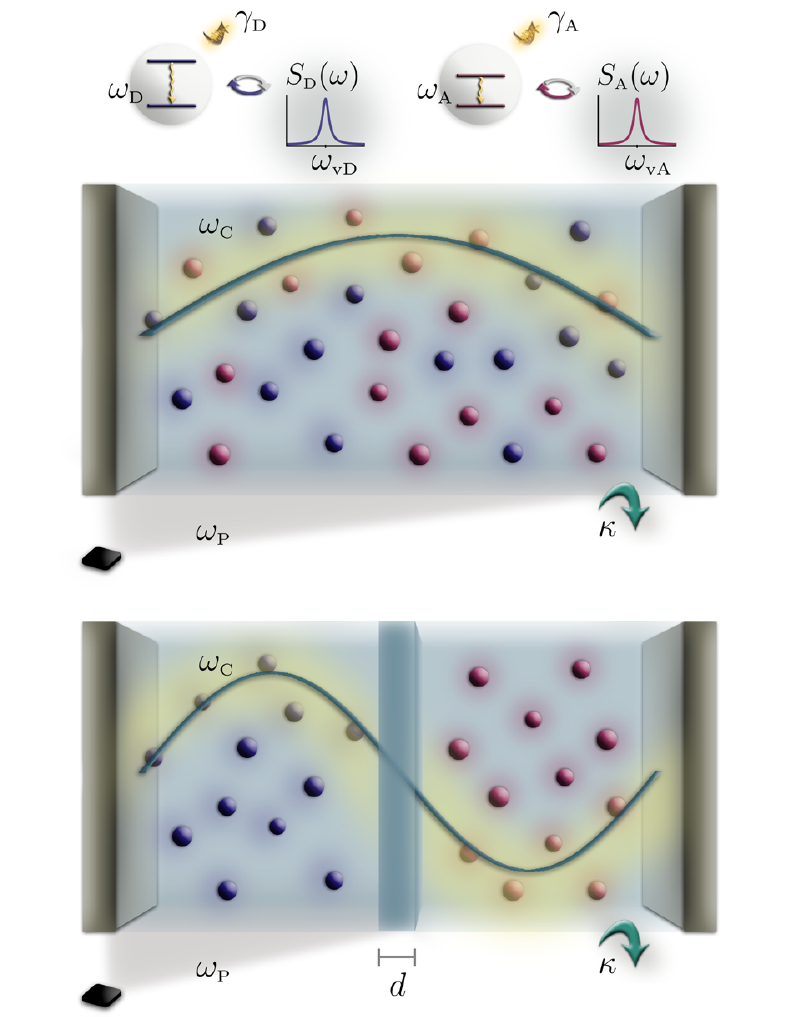}
    \caption{Scheme of the configurations under study, where the molecules
    are placed in the cavity either intermixed (top) or separated by a dividing
    wall (bottom), with the set of parameters characterizing the model. }
    \label{fig:Amodel}
\end{figure}

\autoref{fig:Amodel} sketches our model system. It is composed of
two collections of quantum emitters, $N_{\rm D}$ donors and
$N_{\rm A}$ acceptors, between which the energy transfer is
sought. Both donors and acceptors are treated as two-level systems 
with transition frequencies $\omega_{\sD}$ and
$\omega_{\sA}$, respectively. They are placed, either intermixed
(top panel) or separated by an intermediate wall of width $d$
(bottom panel), inside an optical cavity which supports a single
mode of frequency $\omega_{\sC}$. In the first configuration, which tries to mimic the 
experimental setups analysed in \cite{Coles2014,Ebbesen2016}, the
coupling $g_n$ between the $n$-th molecule and the electromagnetic mode is 
assumed to follow the spatial profile of the fundamental cavity mode. 
In the second scenario, $g_n$ presents a dependence dictated by the second 
cavity mode (note that the coupling strength profile exhibits a
node at the position of the dividing barrier). In this way, we match the experimental 
configuration reported in Ref.~\cite{Ebbesen2017}. The Hamiltonian
describing these two hybrid systems within the rotating wave approximation is given by
\begin{align}
    H_{\rm exc} &= \omega_{\sC} a^\dagger a
    + \sum_{n=1}^{N_{\sD}} \omega_{\sD}\sigma_n^z +\sum_{n=1}^{N_{\sA}} \omega_{\sA}
    \sigma_n^z +
    \nonumber
    \\
& + \sum_{n=1}^{N_{\sD}+N_{\sA}} g_n (a^\dagger \sigma_n +
a\sigma_n^\dagger)
    \ , \label{eq:H}
\end{align}
where $a^\dagger$ and $a$ are the bosonic operators for the cavity
mode, and $\sigma_n^\dagger$ and $\sigma_n$ are the
fermionic operators for the $n$-th molecule ($\sigma_n^z \equiv
[\sigma_n^\dagger, \sigma_n]/2$). Notice that, within this approach, we 
disregard dipole-dipole coupling between molecules as its contribution to exciton transport is negligible when 
the polariton-mediated mechanism is fully operative \cite{Feist2015}.

\begin{figure}[!t]
    \centering
    \includegraphics[width=\linewidth]{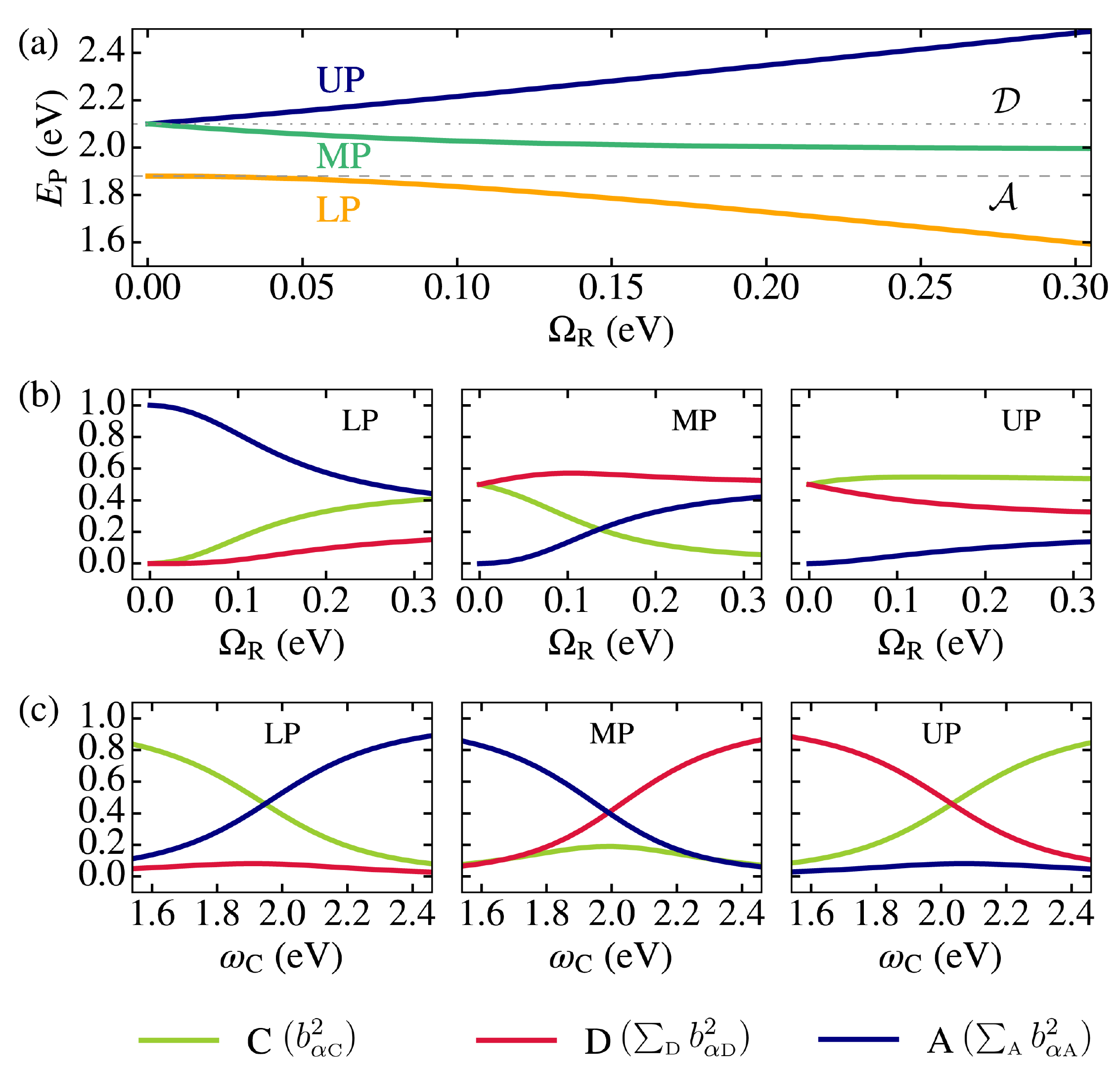}
    \caption{(a) Energies of the UP (blue), MP (green), and LP
     (yellow) as a function of the Rabi frequency $\Omega_{\sR}$.
     Dashed and dash-dotted lines stand for the donor and acceptor
     energies and their associated dark states $\mathcal{D}$ and $\mathcal{A}$.
     (b-c) Coefficients representing the cavity (green), donor (red), and acceptor
      (blue) components of the LP (left), MP (center), and UP (right) as a function of the Rabi
      frequency $\Omega_{\sR}$ for $\omega_{\sC}=2.1$ eV in (b), and the cavity frequency $\omega_{\sC}$ with  
      $\Omega_{\sR}=0.16$ eV in panels (c).}
    \label{fig:Bpolaritons}
\end{figure}

\autoref{fig:Bpolaritons}(a) shows the eigenenergies obtained from
the diagonalization of $H_{\rm exc}$ parametrized in accordance
with the physical magnitudes reported in Ref.~\cite{Ebbesen2017}.
Specifically, in our study, donor and acceptor molecules are
characterized by transition frequencies $\omega_{\sD}=2.1$ eV and
$\omega_{\sA} = 1.88$ eV, respectively, and the cavity mode is
tuned to be at resonance with $\omega_{\sD}$. The eigenenergies are plotted as a
function of the Rabi frequency, $\Omega_{\sR} \equiv
(\sum_{n=1}^{N_{\sD}} g_n^2)^{1/2} $. It is important to notice that the eigenenergies 
only depend on $\Omega_{\sR}$ and not on the particular set of $g_n$. Therefore, 
\autoref{fig:Bpolaritons}(a) is valid for both the intermixed and
physically-separated molecular arrangements. The energies of the upper (UP),
middle (MP), and lower (LP) polaritons are depicted in blue, green
and yellow respectively. Together with these three hybrid
light-matter states, there appear $N_{\sD}-1$ ($N_{\sA}-1$) dark
superpositions of donor (acceptor) states. These constitute the $\mathcal{D}$
and $\mathcal{A}$ subspaces, degenerate at the energies of the bare donor
(grey dash-dotted line) and acceptor (grey dashed line) bare
molecules, respectively.

The cavity and matter components of each polariton are plotted in
\autoref{fig:Bpolaritons} as a function of the Rabi splitting
(panels b) and the cavity frequency
(panels c). The Hopfield coefficients
$b_{\alpha \iota}$ (where $\iota$ = C, D, or A) describes the
content of cavity (C), donor (D), and acceptor (A) of the state
$\alpha$ of the polariton basis (UP, MP and LP). 
For the set of parameters
considered, the UP branch mainly results from the hybridization of cavity and donor
molecules, while the LP branch is composed mostly of acceptor
states mixed with the optical field. This is why UP and LP are
essentially identified with donor and acceptor molecules
respectively, although the presence of other components should not
be overlooked. On the contrary, the MP branch contains a mixture
of states in which both types of molecules have similar weights, which has 
profound implications for long-range energy transfer.

In order to study polariton-mediated energy transfer in the
systems depicted in \autoref{fig:Amodel}, our description must go
beyond Equation~\eqref{eq:H}. First, we need to add a new term in the
Hamiltonian, $H_{\rm P}=\Omega_{\sP} (a^\dagger e^{-i\omega_{\rm
P}t}+ a e^{i\omega_{\rm P}t})$, accounting for the coherent
pumping of the optical cavity by a laser of frequency
$\omega_{\sP}$ and driving strength $\Omega_{\sP}$.  
Within an open quantum system treatment based on the master equation for the 
density matrix, 
we describe the dissipation experienced by molecular excitations due to their
internal vibronic structure by means of the general Bloch-Redfield
approach~\cite{Bloch1957,Redfield1957,Bloch1953}. This requires
the inclusion of spectral densities, $S_n(\omega)$, characterising
the local vibrational reservoir of each molecule $n$. Finally,
radiative losses associated with donors and acceptors (with decay
rates $\gamma_{\sD}$ and $\gamma_{\sA}$ respectively), as well as
cavity losses (with decay rate $\kappa$), are modeled by means of Lindblad
superoperators~\cite{OpenBook,CarmichaelBook}.

\begin{figure*}[htb]
    \centering
    \includegraphics[width=\linewidth]{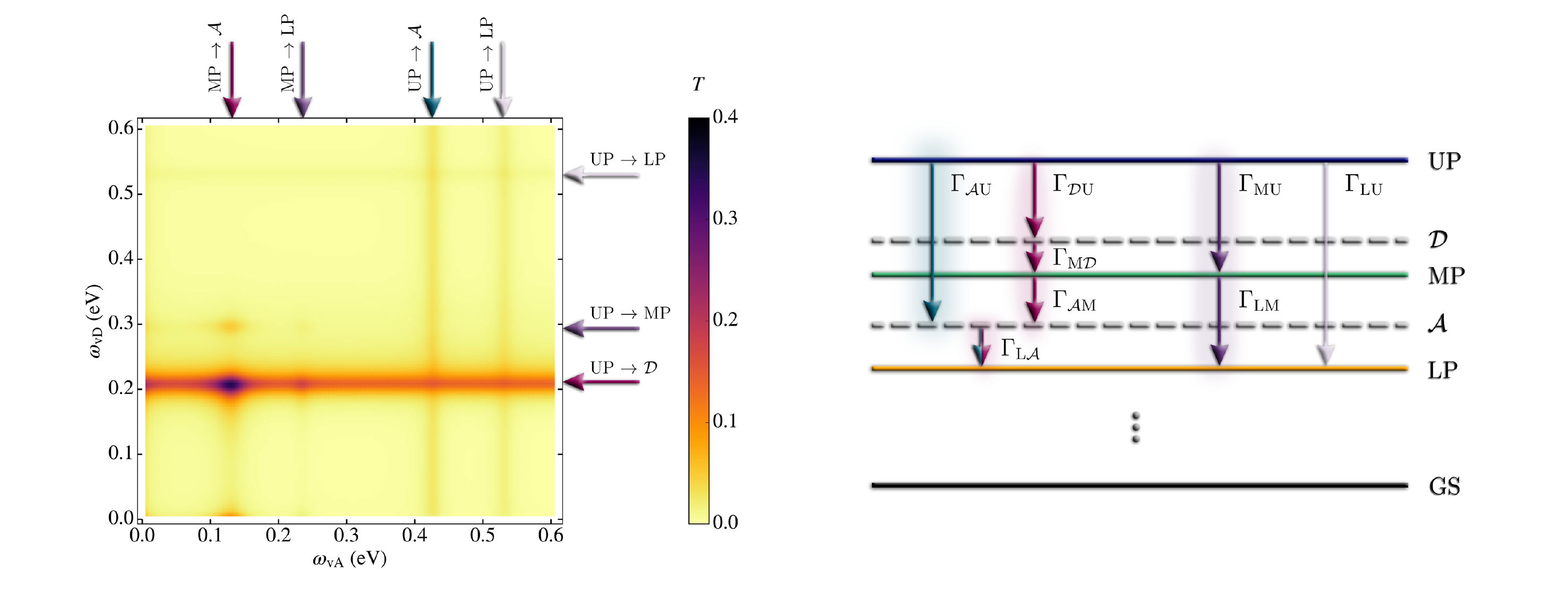}
    \caption{(a) Transfer efficiency, $T$, for $N_{\sD} = 16$ donors and
    $N_{\sA} = 16$ acceptors, coupled such that the Rabi splitting is $\Omega_{\sR} = 0.16$ eV,
    when varying the center frequencies $\omega_{\rm v \sA}$ and $\omega_{\rm v \sD}$ of the
    Lorentzians representing the thermal bath associated with donor and acceptor molecules,
    respectively. The corresponding transitions are marked by arrows on the top and right
    edges. (b) Level scheme of the polaritons (as well as the two sets of dark states $\mathcal{D}$ and $\mathcal{A}$) showing
    various paths of decay, which are labelled with their corresponding rates $\Gamma$.}
    \label{fig:CD_levels}
\end{figure*}

To gain insight into the phenomenon of energy transfer,
we consider a simplified system in our numerical simulations. It consists of
$N_{\sD} = 16$ donors and $N_{\sA} = 16$ acceptors extended over a
length of 100 nm and separated by a wall of $d = 10$ nm. The
molecules are coupled to a cavity mode with a Rabi splitting
$\Omega_{\sR} = 0.16$ eV,  with $g_n$ following the spatial profile of the second cavity mode. The
(non-radiative and radiative) losses of the cavity are set to $\kappa = 0.01$ eV, and the radiative losses of the molecules are
$\gamma_{\sD}, \gamma_{\sA} = 1.3 \ \mu$eV, which are typical values for 
organic molecules. The laser is
tuned to pump the cavity at the UP frequency, so that all the
excitation is placed in this polariton and, consequently, mainly in the
donor molecules. To investigate the role of vibrations in the UP-to-LP transfer of
excitations, we fix all the parameters of the
model except for the spectral densities mimicking the molecular
vibrational reservoirs. Two Lorentzian-like profiles vanishing at
zero frequency are chosen to describe them
\begin{equation}
    S_\iota (\omega) = \gamma_\phi \frac{\omega}{\omega_{\rm v \iota}} \frac{\xi ^2}{(\omega - \omega_{\rm v \iota})^2 + \xi^2} \ ,
\end{equation}
where $\omega_{\rm v \iota}$ are the resonant vibrational
frequencies ($\iota$ = D, A stand for donors and acceptors,
respectively). In accordance with the large dephasing rates present in organic molecules, we take $\gamma_\phi = 0.013$ eV. 
Regarding the linewidth of these vibrational modes, we consider a large value, $\xi = 0.01$ eV, in order to account for 
the effective broadening of the resonances due to cavity losses ($\xi \approx \kappa$).

We theoretically quantify the energy transfer process in a similar way to
experiments, where the light emission at the LP is measured, by defining 
the transfer efficiency, $T$, as 
\begin{equation}
    T = \frac
    { b_{\sL \sC}^2 P_{\sL}}
    {b_{\sL \sC}^2 P_{\sL} + b_{\sM \sC}^2 P_{\sM}  + b_{\sU \sC}^2 P_{\sU}} \ ,
\end{equation}
where $P_\alpha$ is the population of polariton $\alpha$, whose
cavity content is given by coefficient $b_{\alpha \sC}$. Note that
$T$ gives the contrast between the emission from the LP and the
total light leaking from the cavity under coherent pumping.
\autoref{fig:CD_levels}(a) renders $T$ as a function the two
resonant vibrational frequencies, $\omega_{\rm v \sD}$ and
$\omega_{\rm v \sA}$. The relevant energy differences between
eigenstates are indicated in the margins. The contour plot reveals
the conditions for which $T$ is enhanced. We can observe first a
vertical line, which originates from the resonant tuning of
$S_{\rm A}(\omega)$ with the transition from the UP to the $\mathcal{A}$-states. 
This feature cannot be attributed to an energy transfer
process, since it is the acceptor component of the UP that
directly produces it. This enhancement of exciton transport mediated by 
polaritons has already been predicted ~\cite{Feist2015,GonzalezBallestero2015}.
There are also two equivalent high-$T$ lines, one horizontal and one vertical, where
the vibronic frequencies are the same as the energy gap between
the UP and the LP. 

Apart from those, a more pronounced enhancement
of the transfer efficiency takes place when $S_{\rm D}(\omega)$
peaks at the energy gap between the UP and the $\mathcal{D}$-states.
Importantly, $T$ is maximum at a single point along this
horizontal line. This maximum emerges when $\omega_{\rm v \sA}$
also matches the energy difference between the MP and the $\mathcal{A}$-states. 
Therefore, our numerical results reveal that the main pathway that leads to energy
transfer from donor to acceptor molecules carries the population
through the MP thanks to local exciton-vibration interactions.
The mixed composition of the MP, which combines donor and acceptor molecule populations 
in similar proportions, boosts the population transfer. This prevailing path
corresponds to the red route depicted in
\autoref{fig:CD_levels}(b), where the other possible processes are
also shown: the direct decay of the UP into $\mathcal{A}$-states (green) and
the channels involving only polaritons (purple and grey).

Importantly, we have verified that the general picture offered by
\autoref{fig:CD_levels} on the link between energy transfer and
molecular vibrations is not altered, even at the quantitative
level, when donor and acceptor molecules
are intermixed, as long as the Rabi splitting
remains the same. This demonstrates that $\Omega_{\sR}$ is the only key
parameter describing the effect of light-matter coupling in the
process of energy transfer, which is independent of aspects such
as the actual molecular arrangement or the spatial dependence of
the cavity mode.

Our numerical analysis also shows that, for the steady-state solution, terms coupling
the off-diagonal (coherences) and diagonal (populations) elements of the density
matrix can be disregarded in the Bloch-Redfield master equation. Under this approximation
(equivalent to the secular approximation), we can now restrict our attention
to just the transition rates connecting polaritonic and/or dark states,
as depicted in panel (b) of \autoref{fig:CD_levels}. 
This greatly simplifies the numerical treatment, making the theoretical study of systems involving a much
larger number of donor and acceptor molecules feasible. Moreover,
as shown below, we can obtain analytical expressions for the relevant decay rates, 
expressed only in terms of the Hopfield coefficients 
and the vibronic spectral densities.

Depending on the nature of the states involved in the transition,
we can identify three different sets of decay rates. First, those
three connecting two polariton states, $\alpha$ and $\beta$, can be
expressed as
\begin{equation}
\Gamma_{\rm \alpha \beta} = \sum_{\iota=\rm D, A} b_{\beta
{\iota}}^2 b_{\alpha {\iota}}^2 S_{\iota} (\omega_{\beta} -
\omega_{\alpha}) \ , \label{eq:g1}
\end{equation}
where the dependence of $b_{\alpha \iota}$ on the number of
molecules is of the form $b_{\alpha \iota} \sim 1/
\sqrt{N_{\iota}}$, yielding $\Gamma_{\rm \alpha \beta}   \sim 1/
N_{\rm A,D}$. Equation~\eqref{eq:g1} reveals that $\Gamma_{\rm
MU}$, $\Gamma_{\rm LM}$ and $\Gamma_{\rm LU}$ vanish as the number
of molecules increases, and the contribution to the energy
transfer from the purple and grey routes in
\autoref{fig:CD_levels}(b) are negligible in very large systems.

Second, the rates for the decay of the dark subspaces to a polariton state of 
lower energy are given by
\begin{align}
\Gamma_{\alpha \iota' } & = \frac{1}{N_{\iota}} \sum_{\iota}
b_{\alpha \iota}^2 S_{\iota} (\omega_{\iota} - \omega_{\alpha}),
\label{eq:g2}
\end{align}
where $\iota=\rm D$ ($\iota=\rm A$) for $\iota'=\rm \mathcal{D}$
($\iota'=\rm \mathcal{A}$) and $\alpha$ denotes LP for $\iota'=\rm \mathcal{A}$ and LP or MP when $\iota'=\rm \mathcal{D}$. 
Equation~\eqref{eq:g2} indicates that
$\Gamma_{\rm M \mD}$, $\Gamma_{\rm L \mD}$ and $\Gamma_{\rm L \mA}$ present
the same $1/ N_{\rm A,D}$ dependence as the decay rates between
polaritonic states. However, in contrast to decay from polariton states (which decay efficiently
by cavity leakage of their photonic contribution), the competing decay paths due to bare-molecule
radiative and nonradiative decay are typically on the order of nanoseconds for
high-quantum-yield emitters. Consequently, even slow decay from the dark states efficiently
populates the lower-lying polaritons.

Finally, the transition rates from polariton states to dark
subspaces that have lower energies have the form
\begin{align}
    \Gamma_{\iota' \alpha} &
    = \frac{N_{\iota}-1}{N_{\iota}}  \ \sum_{\iota}  b_{\alpha \iota}^2 S_{\iota} (\omega_{\alpha} -
    \omega_{\iota}),\label{eq:g3}
\end{align}
where the $N_\iota -1$ term reflects the (large) number of dark states to
which polaritons can decay. Equation~\eqref{eq:g3} yields
$\Gamma_{\rm \mD U}\sim (N_{\sD}-1)/N_{\sD}$ and $\Gamma_{\rm \mA U},
\Gamma_{\rm \mA M} \sim (N_{\sA}-1)/N_{\sA}$. Thus, these three decay
rates do not decrease as the number of molecules increases, as the other six transition 
rates do. This gives analytical support to our Bloch-Redfield numerical results that showed 
the prevalence of the red route in the decay of the excitation from the UP to the LP.   

Our analytical results not only explain the numerical findings previously
discussed but also serve as a guideline to optimize the long-range energy transfer
mediated by strong coupling. In order to enhance the transfer
efficiency, the vibration-driven decay $\Gamma_{\rm \mD U}$ ($\Gamma_{\rm \mA
M}$) from the upper (middle) polariton to the $\mathcal{D}$ ($\mathcal{A}$) dark
states has to be comparable or faster than its decay $b_{\sU \sC}^2 \kappa$
($b_{\sM \sC}^2 \kappa$) due to cavity losses. This can be achieved most
straightforwardly by using cavities with very low losses. However, for given cavity
losses, optimization would rely on minimizing the cavity component of the upper and middle
polaritons while maximizing the donor (acceptor) component of UP (MP). Both
these conditions favor low cavity frequencies, see \autoref{fig:Bpolaritons}(c).
In addition, the vibration-driven decay can be enhanced by bringing the energy
detuning between UP and $\mathcal{D}$ (MP and $\mathcal{A}$) into resonance with
the main vibronic frequency of the donor (acceptor) molecules. Our model
envisages that transfer efficiencies close to $100\%$ can be reached under the
conditions $\Gamma_{\rm \mD U} \gg b_{\sU \sC}^2 \kappa$ and $\Gamma_{\rm \mA M}
\gg b_{\sM \sC}^2 \kappa$.

To conclude, we have presented both a numerical treatment based on the Bloch-Redfield formalism and an 
analytical approach to underpin the physics of the phenomenon of long-range energy transfer mediated by collective strong coupling. 
We have demonstrated the key role played by the delocalized character of the middle polariton in this process as it enables the vibrations 
to transfer the excitation from donor to acceptor molecules. Importantly, this non-local energy transfer is dominated by the 
Rabi frequency and do not depend on the particular arrangement of the molecules inside the cavity or the 
electromagnetic mode spatial profile. Therefore, as long as collective 
strong coupling is achieved, our theoretical results predict that there is no limit in the physical separation attainable 
between donor and acceptor molecules. Not only we have been able to unveil the physical mechanism behind vibration-driven 
long-range energy transfer, but our analytical approach has allowed us to deliver specific recipes to optimize the phenomenon.

We thank J.~del Pino for fruitful discussions.
This work has been funded by the European Research Council under
Grant Agreements ERC-2011-AdG 290981 and ERC-2016-STG-714870, the
EU Seventh Framework Programme (FP7-PEOPLE-2013-CIG-630996 and
FP7-PEOPLE-2013-CIG-618229), and the Spanish MINECO under
contracts MAT2014-53432-C5-5-R and FIS2015-64951-R, as well as
through the ``Mar\'ia de Maeztu'' programme for Units of
Excellence in R\&D (MDM-2014-0377).

%\bibliography{bibTransfer} % Tell bibtex which .bib file to use (this one is some example file in TexLive's file tree)
%\bibliographystyle{apsrev4-1} % Tell bibtex which bibliography style to use


\end{document}